\documentstyle[12pt,a4wide]{article} 
\title{{\small IMI Bank -- PDG Internal Report 3/98 - Available also at www.damianobrigo.it}\\
{\Large \bf On three filtering problems\\ arising in mathematical finance}
\thanks{This work was developed while the first named author was working
at the Risk Management department of Cariplo Bank. A related paper
appeared later on in: {\em Insurance. Mathematics and Economics}, 
22(1) (1998) pp. 53-64.}} 
\author{
Damiano Brigo \\
Product Development Group \\
IMI Bank, San Paolo IMI Group \\
Corso Matteotti 6\\
  20121 Milano, Italy \\
  Fax: 39 02 7601 9324  \\ 
E-mail: brigo@bimimi.it
  \and 
  Bernard Hanzon \\
Dept. Econometrics \\ 
Free University Amsterdam \\
De Boelelaan 1105, 1081 HV \\ 
Amsterdam, The Netherlands\\
Tel: +31-20-4446017\\
E-mail: bhanzon@econ.vu.nl
}
\date{}
\begin{document}

\newcommand{\xhat}{\widehat{X}}
\newcommand{\yhat}{\widehat{Y}}
\newcommand{\vsp}{\vspace{1cm}} 
\newcommand{\hsp}{\hspace{0.5cm}} 
\newcommand{\half}{\frac{1}{2}}
\newtheorem{theorem}{Theorem}[section] 
\newtheorem{proposition}[theorem]{Proposition} 
\newtheorem{lemma}[theorem]{Lemma} \newtheorem{corollary}[theorem]{Corollary} 
\newtheorem{definition}[theorem]{Definition}  

\maketitle
\thispagestyle{empty}

\begin{abstract}
Three situations in which filtering theory
is used in mathematical finance are illustrated at different 
levels of detail.
The three problems originate from the following different works:
\begin{itemize}
\item[1)] On estimating the stochastic volatility model from
    observed bilateral exchange rate news, by R. Mahieu, and 
    P. Schotman;
\item[2)] A state space approach to estimate multi-factors CIR models 
   of the term structure of interest rates, by A.L.J. Geyer, and
   S. Pichler;
\item[3)] Risk--minimizing hedging strategies under partial observation 
   in pricing financial derivatives, by P. Fischer, E. Platen, and 
   W. J. Runggaldier;
\end{itemize}
In the first problem we propose to use
a recent nonlinear filtering technique based on
geometry to estimate the volatility time series from observed
bilateral exchange rates. The model used here is the stochastic
volatility model. The filters that we propose are known as 
projection filters, and a brief derivation of such filters is given.
The second problem is introduced in detail, and a possible use of 
different filtering techniques is hinted at. 
In fact the filters used for this problem in 2) and part of the 
literature can be interpreted as projection filters and we will 
make some remarks on how more general and possibly more suitable
projection filters can be constructed. 
The third problem is only presented shortly. 
\end{abstract}


\section{Introduction}
The filtering problem consists of estimating a stochastic
process $X_t$ representing an unobserved signal,
on the basis of the past and present observations
$\{Y_s: \ \ 0 \le s \le t\}$ of a related measurement process $Y$.
The information given by the measurement process
up to time $t$ is represented by the $\sigma$-algebra ${\cal Y}_t$
generated by $\{ Y_s:\  0 \le s \le t\}$.
For a quick introduction to the filtering problem see Davis and Marcus
(1981) \cite{davis81b}. For a more complete treatment see Liptser and 
Shiryayev (1978) \cite{liptser77a} from a mathematical point of view or 
Jazwinski (1970) \cite{jazwinski70a} for a more applied perspective.  
The solution of the filtering problem is the conditional density 
$p_{X_t|{\cal Y}_t}$ of the signal $X_t$ given the 
observations ${\cal Y}_t$.
Such a solution in general takes its values in an infinite
dimensional function space in an essential way, as proven in 
Chaleyat-Maurel and Michel (1984) \cite{ChalMaur}. 
As a consequence, in general the filter cannot be implemented by
an algorithm which updates only a finite number of parameters.
This means that there can be no finite-memory computer 
implementation.
An important exception is the
linear-Gaussian case, where the solution $p_{X_t|{\cal Y}_t}$
is Gaussian at all time instants, and as such can be parameterized
by mean and variance. This is the well known Kalman filter.

In the present paper we investigate three possible roles of
filtering theory in mathematical finance. 

The first problem concerns the stochastic volatility models.
In recent applications, time varying volatility of financial time series
has been modelled according to the stochastic volatility model, where
the variance is considered to be a stochastic process representing
an unobserved component. There are several reasons for which
such a model represents a convenient choice: among them, the fact that
such models are related to 
the type of diffusion processes one encounters in finance (asset pricing
theory, see Melino and Turnbull (1990) \cite{MeliTurn}).
Once the type of model is chosen, there are two problems to be solved:

\begin{itemize}

\item[i)] estimate the model parameters on the basis of the 
observed bilateral exchange rates;

\item[ii)] estimate the volatility time series on the basis of the 
observed bilateral exchange rates.

\end{itemize}

We 
%shortly report and comment on the estimation procedure suggested in
%Mahieu and Schotman (1994) \cite{MahiScho} for both i) and ii), and 
develop point ii) by suggesting a different approach
based on the projection filter of Brigo, Hanzon and Le Gland
(1995) \cite{BrHaLe}, (1997) \cite{Bernoulli}.

We continue by considering as a second problem the state space approach 
of Geyer and Pichler (1996) \cite{geyer96a}. Such an approach  
is used to estimate and test multi-factors Cox-Ingersoll-Ross (CIR) 
models of the 
term structure of interest rates. We concentrate on the estimation
procedure. We report the quasi-maximum-likelihood approach
combined with a Kalman filter as suggested by Geyer and Pichler,
and we also hint at a possible completely Bayesian approach 
which is sometimes used in system identification. 

This state--space approach is convenient for several reasons.  
The model is estimated, as in the classical cross--section
approach, from observations of yields. However, in the state--space approach 
yields are modelled by taking in account some noise. In this way,
market imperfections and deviations from the true model are
taken in account. Other advantages are listed in the section
of the paper devoted to this approach, and are presented in
larger detail in Geyer and Pichler (1996) \cite{geyer96a}.

The third problem presented concerns
risk--minimizing hedging strategies under partial observation 
in pricing financial derivatives, and is reported as 
from Fischer, Platen and Runggaldier (1996) \cite{fischer96a}.
This result is reported and commented in a concise fashion,
since it has been thoroughly developed by the authors.
It is an excellent example of how filtering theory can fit nicely
the mathematical-finance setup, and such examples are rare 
in the literature.

%
%
%      P R O B L E M      O N E
%
%

\section{On estimating the stochastic volatility from
       observed bilateral exchange rate news}

\subsection{Introduction}
The main problem econometricians face when dealing with a stochastic 
volatility model is the intractability of the likelihood function. 
In fact, the function turns out to involve a multiple integration, due 
to the unobserved stochastic variance. One can try to remedy this situation
by using a quasi maximum likelihood (QML) method. Another possible
remedy is the method of moments estimation (MME). Unfortunately,
it has become clear that both methods are not always reliable
(see Jacquier, Polson and Rossi (1994) \cite{JaPoRo} and 
Andersen (1994) \cite{Andersen}). In Mahieu and Schotman
(1997) \cite{MahiScho} a study 
of several possible estimation techniques is presented, and once the 
model has been estimated a Kalman smoother is applied to estimate
the volatility time series. In order to do this, the model is transformed
into a linear one and
approximations are made to express the new additive noise,
whose exact distribution is a log chi-squared.
Some possibilities include the approximation of such new noise
by a Gaussian of mean $-1.27$ and 
variance $\pi^2/2$ (QML). Another possible choice is to approximate 
the new noise via a mixture of Gaussian densities
which should approximate the log chi-squared distribution
and other possible noise-distributions in a rather satisfactory way. 
In  Mahieu and Schotman (1997) \cite{MahiScho} 
an application of all the mentioned techniques 
to financial data is considered, and conclusions are drawn. 
In the following we suggest a different
possible approach to the estimation of the volatility time series
from observed bilateral exchange rates.
Once the model has been estimated,
instead of transforming the original (nonlinear) stochastic volatility
model into a linear one and approximating the log chi--squared noise, 
we keep the original {\em nonlinear} system with Gaussian white noise and we
propose to adopt nonlinear filtering techniques in order to estimate the 
volatility.
The nonlinear filters we use are the projection filters, which
were defined and investigated in continuous time in
Hanzon (1987) \cite{Lanc}, Hanzon and Hut (1991) \cite{H-H}, 
Brigo (1995) \cite{brigo95e}, (1996) \cite{brigo96b}, \cite{BrigoPhD},
and Brigo, Hanzon and Le Gland (1995) \cite{BrHaLe}, (1997) \cite{Bernoulli}. 
In this paper we give a short derivation of the projection filter
in discrete time, and
we apply the theory for discrete time projection filters
to the stochastic volatility model.
%Moreover, given the connections with asset price theory in finance
%(see for example Mahieu and Schotman (1997) \cite{MahiScho})
%and the use of diffusion processes in that theory, 
%we can consider continuous time volatility models. 
%In such continuous time transpositions of the stochastic
%volatility models the volatility can be estimated exactly
%from the quadratic variation of the bilateral exchange rates.
%This result is based on typical properties of continuous time
%models, so that if a discrete time model is given, it is essential
%to study the filtering problem in discrete time (as we do)
%rather than translating the model in continuous time.

In general, our method features the advantage of fully taking in account
the nonlinear nature of the model adopted. We do not transform the model,
so that, once it has been estimated, the only approximation involved 
in the estimation of the volatility time series is in the filtering 
technique adopted. In a near future, we plan to analyze the quality of such 
approximation by means
of auxiliary quantities associated to the projection filter.

\subsection{Finite dimensional approximation via minimization
of the Kullback--Leibler information} \label{FDAKLI}
In this section we introduce briefly the Kullback-Leibler information
and we explain its importance for our problem. 
Suppose we are given the space $H$ of all the densities 
of probability measures on the real line equipped with its Borel
field, which are absolutely continuous w.r.t. the 
Lebesgue measure. Then define
\begin{equation}
D(p_1,p_2) := E_{p_1} \{\log p_1 - \log p_2 \} \ge 0, \;\;\;p_1,p_2\in H,
\end{equation}
where in general
\begin{displaymath}
   E_{p} \{\phi\} = \int \phi(x) p(x) dx, \;\;\; p\in H.
\end{displaymath}
The above quantity is the well-known Kullback-Leibler information (KLI).
Its non-negativity follows from the Jensen inequality.
It gives a measure of how much the density $p_2$ is displaced w.r.t.
the density $p_1$. We remark the important fact that $D$ is not a 
distance: in order to be a metric, it should be symmetric and satisfy
the triangular inequality, which is not the case. 
However, the KLI features many properties of a distance in a 
generalized geometric setting (see for instance Amari (1985) 
\cite{Amari-book}). 
For example, it is well-known that the KLI is infinitesimally equivalent 
to the Fisher information metric around every point of a finite--dimensional 
manifold of densities such as $EM(c)$ defined below.
Consider a finite dimensional manifold of
exponential probability densities such as
\begin{eqnarray}
EM(c) &=& \{p(\cdot,\theta) : \theta \in \Theta \subset I\!\!R^m\},
\ \ \Theta \mbox{ open in } I\!\!R^m,
 \\ \nonumber
p(\cdot,\theta) &=& \exp[ \theta_1 c_1(\cdot) + ... + \theta_m c_m(\cdot) -\psi(\theta)],
\end{eqnarray}
expressed w.r.t the expectation parameters 
$\eta$ defined by
\begin{eqnarray}
\eta_i(\theta) = E_{p(\cdot,\theta)}\{c_i\} = \partial_{\theta_i} \psi(\theta),
\,\,\,\,i=1,..,m
\end{eqnarray}
(see for example Brigo, Hanzon and Le Gland (1997) \cite{BrHaLe} 
for more details).
We define $p(x;\eta(\theta)):= p(x,\theta)$ (the semicolon identifies
the parameterization).
Now suppose we are given a density $p\in H$, and we want to
approximate it by a density of the finite dimensional manifold $EM(c)$.
It seems then reasonable to find a density $p(\cdot,\theta)$ in $EM(c)$ 
which minimizes the Kullback Leibler information $D(p,.)$.
Compute
\begin{eqnarray*}
\min_{\theta} D(p,p(\cdot,\theta)) &=&
\min_{\theta} \{ E_p [\log p - \log p(\cdot,\theta)] \} \\ 
&=& E_p \log p - \max_{\theta} \{\theta_1 E_p c_1 + ... +
\theta_m E_p c_m - \psi(\theta)\} \\
&=& E_p \log p -\max_{\theta} V(\theta),\\ 
V(\theta)&:=& \theta_1 E_p c_1 + ... + \theta_m E_p c_m - \psi(\theta).
\end{eqnarray*}
It follows immediately that a necessary condition for the minimum to be
attained at $\theta^\ast$ is
$\partial_{\theta_i} V(\theta^\ast) = 0,\,\,\,\, i=1,...,m$
which yields
\begin{eqnarray*}
E_p c_i - \partial_{\theta_i} \psi(\theta^\ast) = E_p c_i -
E_{p(\theta^\ast)} c_i = 0, \ \ \ \ i=1,..,m
\end{eqnarray*}
i.e. $E_p c_i = \eta_i(\theta^\ast), \ \ i=1,..,m$.
This last result indicates that according to the Kullback Leibler
information, the best approximation of $p$ in the manifold $EM(c)$
is given by the density of $EM(c)$ which shares the same $c_i$
expectations ($c_i$-moments) as the given density $p$. This means that
in order
to approximate $p$ we only need its $c_i$ moments, $i=1,2,..,m$.
%A somewhat analogous result was found in continuous time
%with the Fisher metric and the Hellinger distance 
%(see Brigo, Hanzon and Le Gland (1997) \cite{BrHaLe}).

One can look at the problem from the opposite point of view. Suppose 
we decide to approximate the density $p$ by taking in account only
its $m$ $c_i$--moments. It can be proved (see 
Kagan, Linnik, and Rao (1973) \cite{KaLiRa}, Theorem
13.2.1) that the maximum entropy distribution which shares the
$c$--moments with the given $p$ belongs to the family $EM(c)$.

Summarizing:  If we decide to approximate by using $c$--moments,
then  entropy analysis supplies arguments to use the family $EM(c)$;
and if we decide to use the
approximating family $EM(c)$, Kullback--Leibler
says that the "closest" approximating density in $EM(c)$ shares
the $c$--moments with the given density.
\subsection{The stochastic volatility model}

Let $\{S_t, \ t \in T\}, \;\; T=\{0,1,2,3,...\}$ be a stochastic
sequence
describing bilateral exchange rates in time,
and define
$Y_t := \log S_{t+1} - \log S_{t}, \;\;t \in T$.
Assuming that the change $Y_t$ of $\log S_t$ is unpredictable, the standard
stochastic (logarithmic autoregressive) volatility model (SVM) is given by
\begin{eqnarray} \label{SVM}
X_{t+1} &=&  \rho X_t + \sigma W_{t+1},  \\ \nonumber
Y_t &=&  \exp(\frac{X_t+\gamma}{2}) V_t, 
\end{eqnarray}
where $\{W_s,\ \ s \in T\}$ and $\{V_s,\ \ s\in T\}$ are independent standard
Gaussian white noise processes and $\rho,\sigma,\gamma$ are real constants.
Usually the initial condition $X_0$ features a non informative 
density $p_{X_0}$.
In such models the exchange rate features a fat tailed distribution
due to the mixing of $V_t$ and $\exp[(X_t+\gamma)/2]$.
Consider the following nonlinear filtering
problem:

{\em Estimate the stochastic volatility time series}
$\exp[(X_t+\gamma)/2]$ {\em at time $t$ from the following observations}
\begin{equation}
\nonumber
Y_0^t:= \{Y_s, s\in T, s\le t\}
\end{equation}
{\em of the changes in the logarithms of the bilateral exchange rates
up to time $t$}.

The general solution of such a problem consists of the conditional
probability density $p_{X_t|Y_0^t}$, whose knowledge allows one to
compute, among other estimates, the minimum mean square error estimate 
$E\{\exp[(X_t+\gamma)/2]|Y_0^t\}$ of the stochastic volatility.
Such conditional densities obey the following Bayes formula:
\begin{eqnarray}
p_{X_{t+1}|Y_0^{t+1}}(x) &=&
\frac{p_{Y_{t+1}|X_{t+1}}(Y_{T+1};x) \int_{-\infty}^{+\infty}
p_{X_{t+1}|X_t}(x;u) p_{X_t|Y_0^t}(u)\ \  du}
{N(Y_{t+1})}, \\ \\ \nonumber
N(y) &:=& \int_{-\infty}^{+\infty} p_{Y_{t+1}|X_{t+1}}(y;\xi)
\int_{-\infty}^{+\infty}
p_{X_{t+1}|X_t}(\xi;u) p_{X_t|Y_0^t}(u)\ \  du \ \  d\xi.
\end{eqnarray}
From the structure of the processes $X_t$ and $Y_t$ and from the
assumptions on the noises $V_t$ and $W_t$ it follows immediately that
$p_{Y_t|X_t}(y;x) = p_{{\cal N}(0,\exp(x+\gamma))}(y)$ and
$p_{X_{t+1}|X_t}(x;u)= p_{{\cal N}(x,\sigma^2)}(\rho u)$.
Bayes' formula reads now
\begin{eqnarray}
p_{X_{t+1}|Y_0^{t+1}}(x) &=&
\frac{p_{{\cal N}(0,\exp(x+\gamma))}(y)
\int_{-\infty}^{+\infty} p_{{\cal N}(x,\sigma^2)}(\rho u)
p_{X_t|Y_0^t}(u) \ \ du}
{N(y)}.
\end{eqnarray}
This is the exact solution of our filtering problem. However,
this is very difficult to compute. Assume for example that we
can deal with the numerical integration involved above.
The problem is that in order to obtain the density
at time $t+1$, given the density a time $t$, 
%we need to
%compute the integral in the numerator {\em at every} $x \in %I\!\!R$.
%This is clearly a consequence of the fact that the filter
%is infinite dimensional:
one has to update the given density
point by point in the whole real line.
In the next section we suggest a finite dimensional filter
which approximates the exact filter found in this section.
\subsection{A projection filter for the stochastic volatility
model}
Consider now the family $EM(c)$ of exponential densities 
defined in section (\ref{FDAKLI}). 
More specifically, we take the exponential manifold
$EP(m) := \{p(\cdot,\theta): \theta \in \Theta \subset I\!\!R^m\}$,
with $m$ an even positive integer and
with a linear combination of the monomials $x, x^2,\ldots,x^m$ 
in the exponent:
\begin{equation} \label{exfamily1}
p(x,\theta) = \exp\{\theta_1 x + ... + \theta_m x^m - \psi(\theta)\},
\ \ \theta_m < 0.
\end{equation}
In section (\ref{FDAKLI})
we showed that in order to approximate the density $p=p_{X_t|Y_0^t}$
with a density $p(\cdot,\theta)$ of $EM(c)$,
it suffices to find the density in $EM(c)$ such that the $c_i$-expectations
of $p$ and $p(\cdot,\theta)$ match.
With our specific manifold $EM(c)=EP(m)$, these expectations are exactly the 
first $m$ moments of the exponential density.
Then, in computing the projection filter,
we update only the first  $m$ moments.
Suppose we have computed the projection filter at time $t$ via the
expectation parameters $\eta_1(t),...,\eta_m(t)$. Bayes' formula
yields
\begin{eqnarray*}
\eta_j(t+1)= \frac {\int_{-\infty}^{+\infty} x^j
 p_{{\cal N}(0,\exp(x+\gamma))}(y)
\int_{-\infty}^{+\infty} p_{{\cal N}(x,\sigma^2)}(\rho u)
p(u;\eta(t)) \ \ du \ \ dx}
{\int_{-\infty}^{+\infty}
 p_{{\cal N}(0,\exp(\xi+\gamma))}(y)
 \int_{-\infty}^{+\infty} p_{{\cal N}(\xi,\sigma^2)}(\rho u)
 p(u;\eta(t)) \ \ du \ \ d\xi}, \ \ \ j=1,..,m
\end{eqnarray*}
which permits to update the expectation parameters.
Then the new density $p(\cdot;\eta(t+1))$ may be computed recursively
from the previous one $p(\cdot;\eta(t))$.
If one prefers to avoid normalization at every step, 
one can use the scheme
\begin{eqnarray}
\alpha_j(t+1)&=& \int_{-\infty}^{+\infty} x^j
 p_{{\cal N}(0,\exp(x+\gamma))}(y)
\int_{-\infty}^{+\infty} p_{{\cal N}(x,\sigma^2)}(\rho u)
q(u;\alpha(t)) \ \  du \ \ dx, \ \ \ j=0,...,m, \nonumber \\ \\ 
\eta_i &=& \alpha_i/\alpha_0, \ \ \ i=1,...,m,
\end{eqnarray}
where $q(\cdot;\alpha)$ is the 
unnormalized exponential density of 
the family $\{\exp(\theta_0+\theta_1 x+\ldots+\theta_m x^m);\theta_m<0\}$,
characterized by the unnormalized expectation parameters
$\alpha_i,\ \ i=0,1,..,m$. 
Initially, at $t=0$, one can take $\alpha_0(0) = 1,
\alpha_i(0) = \eta_i(0) \ \ i=1,\ldots,m$.
By expanding this last expression one obtains
\begin{eqnarray} \label{expfsvm}
\alpha_j(t+1) &=& \int_{-\infty}^{+\infty} \{ x^j
 \exp[ -\frac{x+\gamma}{2} - \frac{1}{2 \sigma^2} x^2 -
 \frac{1}{2} y^2 e^{-x-\gamma}] \\ \nonumber
& &\int_{-\infty}^{+\infty}
\exp[-\frac{1}{2\sigma^2}
(-2 \rho x u + \rho^2 u^2)] q(u;\alpha(t)) \  du \ \} dx, \ \ \ j=0,...,m.
\end{eqnarray}
This last equation yields the evolution of the $m+1$ parameters $\alpha$
characterizing the projection filter for $EP(m)$. However, there
are some problems in implementing this equation. Mainly, we need
a way to express the exponential density $p(\cdot;\eta)$ explicitly from
the knowledge of the $\eta$. 
Actually, from the theory of exponential families
(see Brigo (1996) \cite{BrigoPhD}, Chapter 3 and references given therein)
we know that the expectation parameters $\eta$ characterize 
the densities of $EP(m)$, but we do not know a direct way to express  
the densities on the basis of such parameters.
On the contrary, from (\ref{exfamily1}) it is clear that the canonical
parameters $\theta$ permit to express the densities of $EP(m)$ explicitly.
In Brigo (1996) \cite{BrigoPhD} (lemma 3.3.3) we give a recursive formula
for $EP(m)$
which allows one to compute the last expectation parameter
$\eta_m$ and the higher order moments 
$\eta_{m+i}= E_{p(\cdot,\theta)} \{x^{m+i}\}$ for all nonnegative integers $i$,  
on the basis of the canonical parameters $\theta$ and of the first $m-1$ 
expectation parameters $\eta_1,...,\eta_{m-1}$.
Define the matrix $M(\eta)$ as follows:
\begin{eqnarray}
   M_{i,j}(\eta) := \eta_{i+j}, \ \ \ i,j=1,2,\ldots,m.
\end{eqnarray}
It is easy to verify that lemma (3.3.3) of Brigo (1996) \cite{BrigoPhD} 
implies the following formula:
\begin{eqnarray} \label{theta_eta}
   \left[
   \begin{array}{c}
       \theta_1\\
       2 \theta_2 \\
       \vdots \\
       m \theta_m
   \end{array}
   \right]
   \ \ = \ \ -M(\eta)^{-1} \ \
   \left[
   \begin{array}{c}
       2 \eta_1\\
       3 \eta_2 \\
       \vdots \\
       (m+1) \eta_m
   \end{array}
   \right].
\end{eqnarray}
From this last equation it follows that we can recover 
{\em algebraically} the canonical parameters $\theta$ from the 
knowledge of the moments $\eta_1,\ldots,\eta_{2m}$ up to order $2m$. 
Then we can compute the projection filter according to the following
scheme:

\begin{itemize}
\item[(i)] Given the initial density $p(x,\theta(0))=p_{X_0}(x)$,
set $t=0$.

\item[(ii)] Assign $t := t+1$.

\item[(iii)] Compute the first $m$ moments of the new projection 
filter density at time $t$ via the formula
\begin{eqnarray*} 
   \alpha_j(t) &=& \int_{-\infty}^{+\infty} \{ x^j
    \exp[ -\frac{x+\gamma}{2} - \frac{1}{2 \sigma^2} x^2 -
    \frac{1}{2} y^2 e^{-x-\gamma}] \\ \\
   & &\ \ \ \int_{-\infty}^{+\infty}
    \exp[-\frac{1}{2\sigma^2}
    (-2 \rho x u + \rho^2 u^2)] p(u;\theta(t-1)) \  du \ \} dx, 
        \ \ \ j=0,...,m, 
    \\ \\
    \eta_i(t) &=& \alpha_i(t)/\alpha_0(t), \ \ \ i=1,...,m.
\end{eqnarray*}

\item[(iv)] Recover the canonical parameters $\theta(t)$
from the moments $\eta_1(t),\ldots,\eta_{m}(t)$
(What is the best way of doing this is still under
investigation).  

\item[(v)] Estimate the stochastic volatility by evaluating
numerically the integral
\begin{eqnarray*}
    E_{p(\theta(t))}\{\exp(\frac{x+\gamma}{2})\} \ = \
    \int_{-\infty}^{+\infty} \exp(\frac{x+\gamma}{2}) \ 
    p(x,\theta(t)) \ dx.
\end{eqnarray*}

\item[(vi)] Start again from {\bf (ii)}. 

\end{itemize}

A possible problem in applying the above scheme is that for 
the integrals appearing in (iii) and (v) there are apparently no 
closed form expressions while the {\em numerical} integration is a subtle 
problem in this case. One of the difficulties in the numerical
evaluation of the above integrals is that if the filter performs
very well then the resulting density becomes very peaked, so that
special numerical integration techniques are required.  
This problem is currently under investigation.  

A possible heuristic answer to the problem under investigation
in point (iv) is to replace points (iii) and (iv) by the following:

\begin{itemize}
\item[(iii.a)]
Compute the first ${\bf 2m}$ moments of the new projection 
filter density at time $t$ ($j$ and $i$ range now up to $2m$).

\item[(iv.a)] Recover the canonical parameters $\theta(t)$
from the moments $\eta_1(t),\ldots,\eta_{2m}(t)$ by using (\ref{theta_eta}). 

\end{itemize}
For a study of the behaviour of such 
a heuristic procedure, in a slightly different context, and for a 
comparison to several alternatives, including a Newton method,
see Borwein and Huang (1995)  \cite{BorweinHuang95}.
Further investigations into this so called polynomial moment
problem are called for. Better insight into the geometry of the
manifolds $EP(m)$ is likely to be helpful, especially to understand
the behaviour of the various algorithms at the boundary of the
manifold where $\theta_m$ is close to zero.

Concerning the scheme as a whole, difficulties in numerical integration 
in the various steps are still present.
A good performance of the above scheme is not guaranteed and 
it should be tested on simulations. We hope to return to this matter
in future research work.
\section{A state space approach to estimate CIR models 
of the term structure of interest rates}

We consider one of the most popular models of the term-structure 
of interest rates: the multi-factor Cox-Ingersoll-Ross (CIR) model. 
In this model one assumes the instantaneous spot interest-rate $r$ to 
be the sum of $K$ factors $X$ which follow a square-root process under the 
objective probability measure ${\bf P}$:
\begin{eqnarray} \label{cir}
 r_t = X_t^1 + ... + X_t^K, \ \ 
d X_t^j = k_j(\theta_j - X_t^j)dt + \sigma_j \sqrt{X_t^j} d W_t^j \ ,
\  \ j=1,\ldots,K \ . 
\end{eqnarray}
% \mbox{Typically, a Money Account follows} \ \ dB_t = r_t B_t dt .

Let $\{{\cal F}_t, \ \ t \ge 0 \}$ be the filtration representing
the information available through time.
With some reasonable requirements on the parameters $k,\theta$ and 
$\sigma$,
this model yields an almost surely positive spot-rate $r_t$ for
all $t \ge 0$. This is generally considered as one of the main
advantages of the CIR model.
The term structure is expressed by specifying the {\em price} $P_t(T)$ 
at any time $t$ for a $bond$ which pays $1$ at the maturity time $t+T$.
%when the spot-rate process is $r$.
In order to be able to price such bonds and specify the 
term structure of interest rates,
one needs to specify the attitude towards risk.
This is done by specifying the so-called equivalent martingale measure
${\bf Q}$ or risk neutral measure. For simplicity, this measure is taken 
of a form such that under ${\bf Q}$ the factors $X$ still follow 
a square root process of the CIR type:
\begin{displaymath}
 \frac{d{\bf Q}}{d {\bf P}}|{\cal F}_t = 
\exp\left\{ - \sum_{j=1}^{K} \left[
     \frac{\lambda_j^2}{2 \sigma_j^2} \int_0^t X^j_s \ ds
     + \frac{\lambda_j}{ \sigma_j} \int_0^t \sqrt{X^j_s} \ d 
     W^j_s \right] \ 
\right\} 
\end{displaymath}
Under the risk-neutral measure ${\bf Q}$ the factors follow 
the equation
\begin{eqnarray*}
d X_t^j = [k_j\ \theta_j - (k_j + \lambda_j) X_t^j]dt 
+ \sigma_j \sqrt{X_t^j} d  \widetilde{W^j_t} \ ,
\  \ j=1,\ldots,K \ ,
\end{eqnarray*}
where $\widetilde{W}$ is a standard Brownian motion under 
the risk-neutral measure ${\bf Q}$.
The attitude towards risk can be tuned by the parameters
$\ \lambda_1, ... , \lambda_K$, the so called 
{\em market prices of risk}. 
Set $\alpha = (\lambda,k,\theta,\sigma)$.
{\em Yields} are given by 

\begin{eqnarray*}
y_t(T,\alpha) &:=& \frac{- \log \ P_t(T)}{T}  
 = -\frac{1}{T}
 \sum_{j=1}^K [ \log \ \phi(\alpha_j,T) 
             - \psi(\alpha_j,T)\ X_t^j ], \\ \\
\phi(\alpha_j,T) 
  &=& \left[ \frac{2 \sqrt{h} \ 
       \exp\{(k_j + \lambda_j + \sqrt{h})T/2\} }
      { 2 \sqrt{h} 
   + (k_j + \lambda_j + \sqrt{h}) (\exp\{T \sqrt{h}\} - 1 )}
     \right]^{2 k_j \theta_j / \sigma_j^2} \ , \\ \\ 
\psi(\alpha_j,T)
  &=& \frac{2 (\exp\{T \sqrt{h}\} - 1 )}
  {2 \sqrt{h} 
   + (k_j + \lambda_j + \sqrt{h}) (\exp\{T \sqrt{h}\} - 1 )}\ ,
\\ \\
h &=& (k_j + \lambda_j)^2 + 2 \sigma_j^2 \ . 
\end{eqnarray*}  
which are affine functions of the factors $X$. This is a second
advantage of the CIR model: it yields an affine term-structure.

Once this type of model has been established, one is confronted with
the task of estimating the model parameters 
$\alpha = (\lambda, k, \theta, \sigma)$ on the basis of 
the available information.
This problem is usually treated in two ways, as explained in
Geyer and Pichler (1996) \cite{geyer96a}.

\begin{itemize}
\item[1)] The cross section approach: One fits the quantities
$y_t(T,\alpha)$ given above to observed yields in different 
periods of time, finding in each period the parameter values 
for which the model yields $y_t(T,\alpha)$ are closest to the
actually observed yields in that period. The main objections
to this approach are that the
parameter estimates in general are not the same in different
periods of time, and the fact that
even if they were the same, the real dynamics of the spot rate 
$r$ need not follow the CIR structure.

\item[2)] The time series approach: One fits the SDE's for the $X$'s 
(usually for only one factor) to observable proxies of $X_i$ 
(e.g. prices of T-bills or money-market rates).  
This approach raises the following objection:
fitting to different proxies usually produces different estimates
for the same parameters, so as to be inconsistent with the no-arbitrage 
conditions. Moreover, this approach does not
use available information coming from observed yields. 
\end{itemize}

The following state space approach answers the above objections by using
both the {\em CIR dynamics} and the observed  {\em yields' cross section}
without the above inconsistencies.

The idea can be described as follows: assume that the observed yields 
differ from the yields $y_t(T,\alpha)$
prescribed by the model by a white noise process whose 
variance $\delta^2$ is a new parameter to be estimated.
This noise process can be viewed as a tool for taking into account 
market imperfections and deviations from the true model. 
Among the possible advantages of the state-space approach 
(over the pure cross-section approach and the time-series approach)
stated by Geyer and Pichler (1996) \cite{geyer96a} we recall
the following:
\begin{itemize}
\item There is no need to rely on proxies for the factors $X$,
contrary to the time-series approach;
\item It is possible to estimate the parameters themselves rather
than non-invertible functions of them;
\item It is possible to estimate the factors $X$ themselves, 
not only the parameters of the model;
\item Measurement errors are taken into account explicitly.
\end{itemize}
Let us formalize the observation process as follows:
$\tau_t$ is the vector of the $n_t$ maturities at time $t$,
$\epsilon$ is a discrete-time white noise process, and $Y$
is the process of observed yields, where the capital letter 
is used to distinguish between actually observed yields $Y$ 
and the yields $y$ of the CIR model.  

\begin{eqnarray} \label{yields}
&& \tau_t:=[T_t^1,...,T_t^{n_t}]^T \ , \ 
\chi_{i}(\alpha,\tau_t) = -\frac{1}{T_t^i}
 \sum_{j=1}^K  \log \phi(\alpha_j,T_t^i) , 
 \ \Psi_{i,j}(\alpha,\tau_t) = \frac{\psi(\alpha_j,T^i_t)}{T^i_t} \nonumber
 \\ \\ \nonumber  
&& Y_t^i := y_t(T_t^i) + \delta_i \epsilon_t^i = 
\chi_i(\alpha,\tau_t) + \Psi_{i,\cdot}(\alpha,\tau_t) X_t + \delta_i \ \epsilon_t,
\ \ i=1..n_t \ .
\end{eqnarray}
In vector form the observation process reads 
$Y_t = \chi(\alpha,\tau_t) + \Psi(\alpha,\tau_t) X_t + 
\mbox{Diag}(\delta_1,..,\delta_{n_t}) \ \epsilon_t $, 
where the dimension $n_t$ of the vector varies over time 
with the number of maturities.

Now there are essentially two main possibilities for introducing
filtering theory in this setup.

\subsection{Completely Bayesian approach}
The first approach is completely Bayesian, and is used 
in system identification. It consists of viewing the parameters
as new state variables in order to reduce the problem to a 
nonlinear filtering problem. Set 
\begin{displaymath}
(X_t^{K+j},X_t^{2K+j}, X_t^{3K+j}, X_t^{4K+j},X_t^{5K+i}) := 
( k_j, \theta_j, \sigma_j, \lambda_j , \delta_i ), \ \
j=1,...,k, \ \ i=1,...,n_t \ .
\end{displaymath}
In such a way, the equations of the system (\ref{cir},\ref{yields}), 
including the new state variables are:
\begin{eqnarray*}
&& d X_t^{K+r} = 0, \ \ \ \ r = 1,\ldots,4K+n_t, \\  \\
&& d X_t^j =  X_t^{K+j}(X_t^{2K+j}  - X_t^j)dt + 
  X_t^{3K+j} \sqrt{X_t^j} d W_t^j \ , \ \ \ \ j=1,\ldots,K,   \\ \\
&& Y_m^1 = -\frac{1}{T_m^1}
 \sum_{j=1}^K
 [ \log \ \phi(X_m^{4K+j},X_m^{K+j},X_m^{3K+j},X_m^{2K+j},T_m^1) \\
&& \hspace{2cm} - 
\psi(X_m^{4K+j},X_m^{K+j},X_m^{3K+j},X_m^{2K+j},T_m^1)X_m^j ] + 
X_m^{5K+1} \ \epsilon_m^1
 \\ 
&&\vdots
\\ 
&& Y_m^{n_m} = -\frac{1}{T_m^{n_m}}
 \sum_{j=1}^K [ \log \ 
\phi(X_m^{4K+j},X_m^{K+j},X_m^{3K+j},X_m^{2K+j},T_m^{n_m}) \\ 
&& \hspace{2cm}- 
\psi(X_m^{4K+j},X_m^{K+j},X_m^{3K+j},X_m^{2K+j},T_m^{n_m})X_m^j ]
 + X_m^{5K+n_m} \ \epsilon_m^{n_m}
\end{eqnarray*}
This is a filtering problem with continuous time state $X$ and
discrete time observations $Y$, as described for example in
Jazwinski (1970) \cite{jazwinski70a}.
Indeed, the unobserved signal is $X$,
and the observation process $Y$ consists of a deterministic
functional of $X$ plus some noise $X \ \epsilon$. Notice that the noise 
is state dependent, since components of the state $X$ appear in front
of the white noise process $\epsilon$.
The above filtering problem is nonlinear, and as such is infinite
dimensional. An approximation of its solution can be considered.
For example, one can use the extended Kalman filter 
(see again Jazwinski (1970) \cite{jazwinski70a}) even though
no general analytical result on the quality of the filter estimates
is available. Justifications of the use of this filter are
usually based on heuristics.

\subsection{Quasi Maximum Likelihood}
This method is based on an approximate computation of the 
likelihood function. 
Consider equations (\ref{cir}) for the factors of the CIR model.
One of the advantages of square root processes like $X$ is that
they yield closed formulas for the mean and the variance of the 
factors themselves. 
This is somewhat helpful in establishing 
approximations, although nonlinearities in (\ref{cir}) imply
that mean and variance are not sufficient to characterize the
probabilistic behaviour of the factors $X$, contrary to the 
linear case. 
Indeed, the factor $X^j$ features a non-central $\chi^2$ transition 
density. % $p_{X^j_{m+1}|X^j_m}$.
Define $\xhat^j_{t|s} = E\{X^j_t|Y_1,\ldots,Y_s\}$
and  $V^{j,j}_{t|s} = E\{(X^j_t -\xhat^j_{t|s})^2|Y_1,\ldots,Y_s\}$ 
for $j=1,\ldots,K$ and for any $0 \le s \le t$.
From the above considerations it follows easily that between two 
observations, for $m \le t < m+1$, 
the {\em prediction} step is given by
\begin{eqnarray} \label{predic}
\xhat^j_{m+1|m} &=& \theta_j [1 - \exp(-k_j)] 
              + \exp(-k_j) \xhat^j_{m|m} ,\nonumber \\ \\ \nonumber
V^{j,j}_{m+1|m} &=& \sigma^2_j \frac{1 - \exp(-k_j)}{k_j} 
         [\theta_j \frac{(1 - \exp(-k_j))}{2} + \exp(-k_j) \xhat^j_{m|m} ] 
+ \exp(-2 k_j) V^{j,j}_{m|m}.
\end{eqnarray}
Notice that even if at a certain time the conditional density 
$p^j_{m|m}$ of $X_m^j$ given $Y_1,\ldots,Y_m$ were Gaussian, i.e.
\begin{displaymath}
p^j_{m|m} \sim {\cal N}(\xhat^j_{m|m},V_{m|m})  ,
\end{displaymath}
the prediction step would lead us out of the Gaussian family: 
\begin{displaymath}
p^j_{m+1|m} \not \sim {\cal N}(\xhat^j_{m+1|m},V^{j,j}_{m+1|m}) \ .
\end{displaymath}
Therefore, $p^j_{m+1|m}$ is not Gaussian and its mean $\xhat^j_{m+1|m}$
and variance $V^{j,j}_{m+1|m}$ are not enough to activate the 
{\em correction step} (Bayes' formula) leading to the conditional
density $p^j_{m+1|m+1}$.
In order to avoid such difficulties, one can {\em replace} 
the real $p^j_{m+1|m}$ by  ${\cal N}(\xhat^j_{m+1|m},V^{j,j}_{m+1|m})$,
i.e. replace the density $p^j_{m+1|m}$ by a Gaussian density
sharing its first two moments.
This is actually what is done in 
Geyer and Pichler \cite{geyer96a}. As we remarked earlier in
Section \ref{FDAKLI}, this amounts to
replacing $p^j_{m+1|m}$ by its best approximation, in the 
Kullback--Leibler sense, of the Gaussian family. Therefore
the approximate filter used here can be interpreted as a
Gaussian projection filter!
By this approximation, it follows that the approximated {\em correction}
at $t = m + 1$, when $Y_{m+1}$ is available, is given by
%\begin{displaymath}
%Y_{m+1} =  
%\chi(\alpha,\tau_{m+1}) + \Psi(\alpha,\tau_{m+1}) X_{m+1} + 
%\mbox{Diag}(\delta_1,..,\delta_{n_{m+1}}) \ \epsilon_{m+1} \ 
%\end{displaymath}
% is available, 
Bayes' formula and can be summarized by
%giving $\xhat_{m+1|m+1},V_{m+1|m+1}$ in terms of
%$\xhat_{m+1|m},V_{m+1|m}, Y_{m+1}$:

\begin{eqnarray} \label{correc}
&& \Delta_m := \mbox{Diag}(\delta_1,..,\delta_{n_m}), 
\nonumber \\  \nonumber \\ \nonumber 
&& \xhat_{m+1|m+1} = \{ \xhat_{m+1|m} + 
     V_{m+1|m} \Psi^T(\alpha,\tau_{m+1}) 
   \left[ 
    \Psi(\alpha,\tau_{m+1})  V_{m+1|m} \Psi(\alpha,\tau_{m+1})^T + 
   \Delta_{m+1}^2 
   \right]^{-1} \\ \\ \nonumber
&& \hspace{3cm}  \left(Y_{m+1} - \chi(\alpha,\tau_{m+1}) 
    - \Psi(\alpha,\tau_{m+1}) 
   \xhat_{m+1|m} \right) \}^{\bf +} \ ,  \\ \nonumber \\ \nonumber
&&  V_{m+1|m+1} = \left[ \Psi^T(\alpha,\tau_{m+1}) 
                 \Delta_{m+1}^{-2}
                  \Psi(\alpha,\tau_{m+1})  + V_{m+1|m}^{-1} \right]^{-1} \ .
\end{eqnarray}
The symbol $\{\ \cdot \ \}^+$ in the above equation denotes the positive
part. It is applied in order to make sure that the approximate conditional
mean $\xhat$ be positive. 
We can now calculate the quasi-likelihood function as follows:  
Set $\beta = (\alpha,\delta)$ and compute
\begin{eqnarray*}
&& p_{Y_1,...,Y_n}(y_1,...,y_n;\beta)  
 =  p_{Y_n-\yhat_{n|n-1}}(y_n - \yhat_{n|n-1};\beta) 
   p_{Y_{n-1}-\yhat_{n-1|n-2}}(y_{n-1} - \yhat_{n-1|n-2};\beta) 
   \cdots \\ \\
 && \hspace{2cm} \cdots p_{Y_1-\yhat_{1}}(y_1 - \yhat_{1};\beta)  \\ \\ 
&& =  p_{{\cal N}(0,\Psi(\alpha,\tau_n) V_{n|n-1} 
   \Psi(\alpha,\tau_n)^T + \Delta_n^2)}(y_n - \yhat_{n|n-1}) \\ \\
&&   p_{{\cal N}(0,\Psi(\alpha,\tau_{n-1}) V_{n-1|n-2} 
        \Psi(\alpha,\tau_{n-1})^T + \Delta_{n-1}^2)}
   (y_{n-1} - \yhat_{n-1|n-2})  
  \cdots 
   p_{{\cal N}(0,\Psi(\alpha,\tau_1) V_1 \Psi(\alpha,\tau_1)^T + \Delta_1^2)}
  (y_1 - \yhat_1) .
\end{eqnarray*}
This function can be computed (and maximized) once we know
$\yhat$ and $V$ for all $\beta$. These quantities can be obtained
for every possible value of $\beta$ from the above recursion
(\ref{predic}, \ref{correc}).
Of course, in practice numerical simulation techniques are required
to maximize the quasi-likelihood.

The two unanswered questions about this approach are:
\begin{itemize}
\item How good is the Kullback-Leibler projection on the 
Gaussian family used after the prediction step?
\item How good is taking $\{ \ \ \}^+$ in the correction?  
\end{itemize}

In order to deal appropriately with the first of these questions 
one can make use of the concept of
{\em projection residual} that was developed for the continuous time 
case in Brigo, Hanzon and Le Gland (1995) \cite{BrHaLe}. 
This concept can actually be used here, because the approximate filter 
used in \cite{geyer96a} has in fact the interpretation of a continuous
time Gaussian Projection Filter for a continuous time signal 
observed in discrete time.
Of course the question about taking $\{ \ \ \}^+$ arises 
because here one works with Gaussian densities. In order to avoid this
problem one could try to work with a class of densities which have their
support on the non-negative real halfline and work out the 
Projection Filter, for the model under investigation here, by using
such a class of densities.

\section{Risk--minimizing hedging strategies under partial observation}
%(P. Fischer, E. Platen, W.J. Runggaldier)
We shortly report the result of Fischer, Platen, and Runggaldier
(1996) \cite{fischer96a}. This is a significant case where
filtering theory fits nicely a mathematical-finance setup. 
A financial market is considered over a time interval $[0,T]$
with a risky asset, whose price is denoted by $S$, and a bond,
whose price is assumed identically equal to one. Under a martingale
measure, we write 
%(P. Fischer, E. Platen, W.J. Runggaldier)
\begin{eqnarray*}
&& B_t = 1, \\ \\
&& dS_t = \sigma_t(Z_t) S_t d W_t \\ 
&& dy_t = A_t S_t dt + D_t  dV_t .
\end{eqnarray*}
Let ${\cal F}_t = \sigma\{S_u, Z_u: \ \ u \le t\}$ be the information
represented by observation of $S$ and $Z$ up to time $t$.
The process $Z$ is a hidden Markov process (representing the state of 
the economy) with transition intensity matrix $\Lambda$.
Let $N_t$ be the number of jumps of $Z$ (number of changes in the
economy) up to time $t$.
The process $y_t$ represent observation of $S_t$ in additive noise,
reflecting the possibility that not all indicated prices are 
actually traded.
Our observation process is denoted by $Y_t := [y_t , \ \ N_t]$.
Denote by ${\cal Y}_t := \sigma\{Y_s,\ s\le t\}$ the information
represented by observation of $S$ and $Z$ up to time $t$.
We assume that $S_T$ is fully observed. 
We consider a contingent claim $H= H(S_T)$ to be priced at all 
$t < T$.
We will consider two cases: full observations 
$\{{\cal F}_t\ : t \ge 0\}$ available, and partial observations
$\{{\cal Y}_t\ : t \ge 0\}$ available.
In both cases we are dealing with an incomplete market,
since there are more sources of randomness than traded risky assets. 
Then perfect hedging with
self-financing portfolios is not possible in general.   
We can still try to determine a mean self financing hedging strategy
that minimizes a risk criterion related to the lack of self-financing. 

We begin by the case with full observations.
The main ingredient is the Kunita - Watanabe decomposition.
We are looking for a strategy $(\xi_t,\eta_t)$ 
($\xi_t$  amount of stock, $\eta_t$ amount of bond) such that

\begin{itemize}
\item[i)] $\xi_t$ is ${\cal F}_t$ predictable,  
          $\eta_t$ is ${\cal F}_t$ adapted,
          and 
\begin{displaymath}
   E\{ \int_0^T |\xi_t|^2 \sigma_t(Z_t)^2 S_t^2 dt \} < \infty  \ .
\end{displaymath}

\item[ii)] $\xi_T \ S_T + \eta_T \ 1 = H$ (final value of
          the strategy equals the claim)

\item[iii)] $\xi_t \ S_t + \eta_t \ 1 
           - \int_0^t \xi_u \ dS_u =: C_t(\xi,\eta)$ (value - gains = 
constant) is a martingale (mean-constant);

\item[iv)] minimizes $E\{(C_T - C_t)^2|{\cal F}_t\}$ for each $t$ 
           (quadratic criterion) among
          all other strategies as in (i), (ii), (iii). 
\end{itemize} 
The solution of this problem was derived by F\"{o}llmer and Schweizer
(1991) \cite{follmer91a}.
 They proved, among other results, that if $H \in L_2({\cal F}_T,Q)$ 
($Q$ is a martingale measure for $S$), then 
\begin{eqnarray*}
\xi_t^\ast = \xi_t^H, \ \ \eta_t^\ast = E\{H|{\cal F}_t\} - 
    \xi_t^\ast S_t, \ \mbox{where} \\ \\
E\{H|{\cal F}_t\} = EH + \int_0^t \xi_u^H dS_u + L_t^H
\end{eqnarray*} 
is the Kunita-Watanabe decomposition
($L$ is a martingale, orthogonal to $S$).

In the case of partial observations, points (i), (iii)  and (iv)
are replaced respectively by

\begin{itemize}
\item[i)] $\xi_t$ is ${\cal Y}_t$ predictable,  
       $\eta_t$ is ${\cal Y}_t$ adapted,
and 
\begin{displaymath}
   E\{ \int_0^T |\xi_t|^2 \sigma_t(Z_t)^2 S_t^2 dt | {\cal Y}_0 \} < \infty  \ .
\end{displaymath}

\item[iii)] $E\{\xi_t \ S_t + \eta_t \ 1 
           - \int_0^t \xi_u \ dS_u | {\cal Y}_t\} 
           =  E\{C_t(\xi,\eta)|{\cal Y}_t\}$
          is a $({\cal Y}, Q)-$martingale;

\item[iv)] minimizes $E\{(C_T - C_t)^2|{\cal Y}_t\}$ among
          all other strategies as in (i), (ii), (iii).
\end{itemize} 
The solution of this second problem was given by 
Schweizer (1994) \cite{schweizer94a}, see also
Di Masi, Platen and Runggaldier (1995) \cite{dimasi95a} .

\begin{eqnarray*}
E\{H|{\cal F}_t\} = EH + \int_0^t \xi_u^H dS_u + L_t^H, \\ \\
\xi_t^{\cal Y} = \frac{E\{\xi_t^H\ \sigma^2_t(Z_t)\ S_t^2|{\cal Y}_t\}}
                      {E\{\sigma^2_t(Z_t)\ S_t^2|{\cal Y}_t\}} , \ \ 
 \eta_t^{\cal Y} = E\{H|{\cal Y}_t\} - 
    \xi_t^{\cal Y} S_t. 
\end{eqnarray*} 

How can one compute $\xi^H$ and $E\{H|{\cal Y}_t\}$ explicitly ? 
The solution of this problem was given by Di Masi, Kabanov and Runggaldier
(1994) \cite{dimasi94a}.
If $H$ has polynomial growth, then
\begin{eqnarray*}
\xi_t^H = \xi_t^H(S_t,Z_t) = \frac{\partial }{\partial S} u_t(S_t,Z_t),
 \ \ E\{H|{\cal Y}_t\} = E\{u_t(S_t,Z_t)|{\cal Y}_t\},
\end{eqnarray*}

where $u_t(x,i) = E\{H|S_t=x, Z_t=i\}$ solves

\begin{eqnarray*}
\partial_t u_t(x,i) + \frac{1}{2} \sigma^2(i) x^2 
 \frac{\partial^2 }{\partial x^2} u_t(x,i) + \sum_j \Lambda_{ij} u_t(x,j) = 0, \ \
      u_T(x,i)=H(x)
\end{eqnarray*}

The ${\cal Y}-$mean self-financing strategy can be computed
via the conditional distribution of the unobserved state $(S_t,Z_t)$
given the observations ${\cal Y}_t$. 
This is the filtering problem treated by Miller and 
Runggaldier (1996) \cite{miller96a}. 

\section{Acknowledgements}
This paper was originally presented at the {\em 
International Workshop on the Interplay between Insurance, Finance
and Control}, held at the 
Mathematical Center of the University of Aarhus 
on February 25 -- March 1, 1997. 
%The expenses for
%the participation of the first named author were covered by the
%Mathematical Center of the University of Aarhus and by Cariplo 
%Bank SpA. 
%It is unusual to mention the funding of a trip in a paper;
%funding of the research is another matter-Bernard
The first named author wishes to thank Aleardo Adotti, head of the Product
Development Group of IMI Bank and Renzo G. Avesani,
head of Risk Management and Research at Cariplo Bank, for encouraging
the prosecution of his research activities even in their most mathematical
aspects.

\end{document}